\documentclass[aps,prd,floats,preprint,nofootinbib]{revtex4}
\usepackage{graphicx}
\usepackage{amsmath}
\usepackage{amsfonts}
\usepackage{amssymb}
\usepackage{multirow}
\usepackage{psfrag}
\usepackage{color}
\begin{document}
\def\beq{\begin{equation}}
\def\eeq{\end{equation}}
\def\bea{\begin{eqnarray}}
\def\eea{\end{eqnarray}}
\def\ben{\begin{enumerate}}
\def\een{\end{enumerate}}
\def\bit{\begin{itemize}}
\def\eit{\end{itemize}}

\def\ie{{\it i.e.}}
\def\etc{{\it etc.}}
\def\eg{{\it e.g.}}
\def\lsim{\mathrel{\raise.3ex\hbox{$<$\kern-.75em\lower1ex\hbox{$\sim$}}}}
\def\gsim{\mathrel{\raise.3ex\hbox{$>$\kern-.75em\lower1ex\hbox{$\sim$}}}}
\def\ifmath#1{\relax\ifmmode #1\else $#1$\fi}
\def\half{\ifmath{{\textstyle{1 \over 2}}}}
\def\threehalf{\ifmath{{\textstyle{3 \over 2}}}}
\def\quarter{\ifmath{{\textstyle{1 \over 4}}}}
\def\eigth{\ifmath{{\textstyle{1\over 8}}}}
\def\sixth{\ifmath{{\textstyle{1 \over 6}}}}
\def\third{\ifmath{{\textstyle{1 \over 3}}}}
\def\twothirds{{\textstyle{2 \over 3}}}
\def\fivethirds{{\textstyle{5 \over 3}}}
\def\fourth{\ifmath{{\textstyle{1\over 4}}}}
\def\chitil{\wt\chi}
\def\fbi{~{\mbox{fb}^{-1}}}
\def\fb{~{\mbox{fb}}}
\def\br{BR}
\def\gev{~{\mbox{GeV}}}
\def\gevc{\mbox{GeV}}
\def\tev{~{\mbox{TeV}}}
\def\calm{\mathcal{M}}
\def\mll{m_{\ell^+\ell^-}}
\def\tanb{\tan\beta}

\def\wtil{\widetilde}
\def\cnone{\wt\chi^0_1}
\def\cnonestar{\wt\chi_1^{0\star}}
\def\cntwo{\wt\chi^0_2}
\def\cnthree{\wt\chi^0_3}
\def\cnfour{\wt\chi^0_4}
\def\snu{\wt\nu}
\def\snul{\wt\nu_L}
\def\msnul{m_{\snul}}
\def\se{\wt e}
\def\smu{\wt\mu}
\def\snu{\wt\nu}
\def\snul{\wt\nu_L}
\def\msnul{m_{\snul}}

\def\snue{\wt\nu_e}
\def\snuel{\wt\nu_{e\,L}}
\def\msnuel{m_{\snul}}

\def\snubar{\ov{\snu}}
\def\msnu{m_{\snu}}

\def\snue{\wt\nu_e}
\def\snuel{\wt\nu_{e\,L}}
\def\msnuel{m_{\snul}}

\def\snubar{\ov{\snu}}
\def\msnu{m_{\snu}}
\def\mcnone{m_{\cnone}}
\def\mcntwo{m_{\cntwo}}
\def\mcnthree{m_{\cnthree}}
\def\mcnfour{m_{\cnfour}}
\def\wt{\widetilde}
\def\anti{\overline}
\def\wh{\widehat}
\def\cpone{\wt \chi^+_1}
\def\cmone{\wt \chi^-_1}
\def\cpmone{\wt \chi^{\pm}_1}
\def\mcpone{m_{\cpone}}
\def\mcpmone{m_{\cpmone}}

\def\staur{\wt \tau_R}
\def\staul{\wt \tau_L}
\def\stau{\wt \tau}
\def\mstaur{m_{\staur}}
\def\stauone{\wt \tau_1}
\def\mstauone{m_{\stauone}}

\def\gl{\wt g}
\def\mgl{m_{\gl}}
\def\stl{{\wt t_L}}
\def\str{{\wt t_R}}
\def\mstl{m_{\stl}}
\def\mstr{m_{\str}}
\def\sbl{{\wt b_L}}
\def\sbr{{\wt b_R}}
\def\msbl{m_{\sbl}}
\def\msbr{m_{\sbr}}
\def\sbot{\wt b}
\def\msbot{m_{\sbot}}
\def\sq{\wt q}
\def\sqbar{\ov{\sq}}
\def\msq{m_{\sq}}
\def\slep{\wt \ell}
\def\slepbar{\ov{\slep}}
\def\mslep{m_{\slep}}
\def\slepl{\wt \ell_L}
\def\mslepl{m_{\slepl}}
\def\slepr{\wt \ell_R}
\def\mslepr{m_{\slepr}}
\def\jet{{\rm jet}}
\def\filt{{\rm filt}}
\def\cut{{\rm cut}}
\def\sub{{\rm sub}}
\def\sig{\mbox{sig}}

\def\fT{f_T}
\def\fpi{f_\pi}
\def\nsub{ n_{\text{sub}}}
\def\core{\text{core}}
\def\cores{\text{cores}}
\def\tbar{\overline{t}}
\def\ubar{\overline{u}}
\def\cbar{\overline{c}}
\def\Tbar{\overline{T}}
\def\bbar{\overline{b}}
\def\afb{A_{FB}^t}
\def\ttbar{t\tbar}
\def\tbart{\tbar t}
\def\fat{\text{fat}}
\def\bbbar{b\bbar}
\def\hc{ \text{h.c.}}
\def\detPhi{\text{det}~\Phi}
\def\e{\varepsilon}
\def\Oone{\mathcal{O}(1)}
\def\M{\mathcal{M}}
\def\cut{\text{cut}}
\def\cj{\text{cj}}
\def\fb{\text{fb}}
\definecolor{darkred}{rgb}{0.7,0.0,0.0}
\definecolor{darkblue}{rgb}{0.0,0.0,0.9}
\definecolor{darkgreen}{rgb}{0.0,0.5,0.0}
\definecolor{brown}{rgb}{0.0,0.0,0.0}
\definecolor{white}{rgb}{1.0,1.0,1.0}

\newcommand{\red}{\color{darkred}}
\newcommand{\blue}{\color{darkblue}}
\newcommand{\brown}{\color{brown}}
\newcommand{\green}{\color{darkgreen}}
\newcommand{\white}{\color{white}}
\newcommand{\excola}{\blue}
\newcommand{\excolb}{\blue}

\newcommand{\gtupi}{ { g_{t u \pi}}}
\newcommand{\gtuh}{ { g_{t u h_t}}}
\newcommand{\gturho}{ { g_{t u \rho}}}

\def\CC{{C\nolinebreak[4]\hspace{-.05em}\raisebox{.4ex}{\tiny\bf ++}}}
\newcommand{ \slashchar }[1]{\setbox0=\hbox{$#1$}   
   \dimen0=\wd0                                     
   \setbox1=\hbox{/} \dimen1=\wd1                   
   \ifdim\dimen0>\dimen1                            
      \rlap{\hbox to \dimen0{\hfil/\hfil}}          
      #1                                            
   \else                                            
      \rlap{\hbox to \dimen1{\hfil$#1$\hfil}}       
      /                                             
   \fi}
\widowpenalty=1000
\clubpenalty=1000
\vspace*{3cm}
\begin{flushright}
UMD-PP-013-006\\
\end{flushright}
\title{New light on $WW$ scattering at the LHC with $W$ jet tagging}
\author{Yanou Cui${}^{a}$ and Zhenyu Han${}^{b}$}

\affiliation{  \small \sl ${}^a$ Maryland Center for Fundamental Physics, Department of Physics, University of Maryland, College Park, MD 20742, USA\\
                          ${}^b$ Institute of Theoretical Science, University of Oregon, Eugene, OR 97403, USA
                          }

\def\thesection{\arabic{section}}
\def\thetable{\arabic{table}}


\begin{abstract}

After the recent discovery of a 125 GeV Higgs-like particle at the Large Hadron Collider (LHC), 
it is crucial to examine its role in unitarizing high energy $W_LW_L$ scattering, which may reveal its possible deviation from a Standard Model Higgs. We perform an updated study on $WW$ scattering in the semileptonic channel at the LHC, improved by the recently developed $W$ jet tagging method. The resultant statistical significance of a Strongly-Interacting Light Higgs (SILH) model is about $20\%$ larger than that based on the conventionally ``gold-plated'' dileptonic channel, while $200\%$ more signal events are retained. 
The excellent sensitivity to the anomalous Higgs-$W$ boson coupling makes semileptonic $WW$ scattering an important complement to precision measurements at the Higgs resonance.

\end{abstract}

\maketitle
\thispagestyle{empty}

\pagenumbering{arabic}


\section{introduction}
\label{sec:introduction}
Longitudinal $WW$ scattering at high energies directly probes the mechanism of electroweak symmetry breaking (EWSB). In particular, with unitarity as a guideline this process has been seen as a promising place to search for new physics if EWSB involves strong dynamics such as in the Higgsless models. The recent exciting discovery of a Higgs-like particle at the LHC \cite{higgs_discovery} does not eliminate the importance of $W_LW_L$ scattering. This is in part because the measurements \cite{new_measurements} of the cross sections of the Higgs-like particle still bear significant uncertainties as for now, which allows the possibility that the observed resonance is a Higgs imposter or a dilaton/radion \cite{imposters}, while EWSB follows the pattern of a Higgsless model. Of course a more naturally favored scenario in light of the recent data is that the observed new particle is indeed mostly a Standard Model (SM) Higgs boson \cite{newfit}. In this case it is important and intriguing to explore possible deviations in Higgs properties from the SM predictions, which can originate from an extended Higgs sector or other new physics connected to the Higgs \cite{ww}. $W_LW_L$ scattering is an important place to look for such deviations, since a key role of the SM Higgs is to unitarize this process at high energies.

   Most of the efforts exploring beyond-the-SM properties of the Higgs-like new particle have been focusing on extracting \textit{on-shell} couplings from measurements at the Higgs resonance. Another prospect not as well stressed is that: processes involving an \textit{intermediate off-shell} Higgs may provide complementary information to what we can learn from the on-shell measurements. $W_LW_L$ scattering as the focus of this paper, can be seen as such an example. First, in regard to measuring the Higgs-$W$ boson coupling, unlike precision measurements at the Higgs pole, $W_LW_L$ scattering is insensitive to uncertainties from invisible or exotic Higgs decays. Furthermore, with enough statistics $W_LW_L$ scattering may resolve different models that can be nearly degenerate at the Higgs resonance. For instance, a two-Higgs-doublet model and a composite Higgs model \cite{Giudice:2007fh} both may include a 125 GeV scalar with a coupling to the $W$ boson deviated from that of the SM Higgs by a similar amount. On the other hand, the two models may have very different behaviors in high energy $W_LW_L$ scattering which captures the sum-over effect from all intermediate states responsible for EWSB.

    As a case study, we explore a generic class of composite Higgs models, the Strongly Interacting Light Higgs (SILH) models \cite{Giudice:2007fh}. The signal of such models is defined as the event excess in $W_LW_L$ scattering over the SM prediction with a light Higgs. We investigate the LHC sensitivity to the excess in the semileptonic decay channel, using the recently developed $W$-jet tagging technique \cite{wtag}. The signal of the SILH model can be approximated by that of the Higgsless model, scaled down by a factor $\sim (c_H\xi)^2$, which quantifies the deviations of the Higgs-weak boson couplings from the SM \cite{Giudice:2007fh, Han:2009em}. This allows us to focus on the limiting case of the Higgsless model and then infer the sensitivities for the SILH models using the scaling relation. The Higgsless model signal is obtained by turning off diagrams involving the Higgs boson in the SM, without introducing heavy $WW$ resonances \cite{Alboteanu:2008my}. The existence of such resonances is implied by the breakdown of perturbative unitarity, which typically enhances the signal strength. Therefore, the sensitivities derived in our study can be seen as a conservative estimate. Meanwhile, we emphasize that if these resonances weakly couples to the fermions, it is essential to search for them in the $WW$ scattering signal.

     Compared to the dileptonic channel, the well-known challenge for the semileptonic $W_LW_L$ scattering is the contamination from QCD backgrounds. However, the semileptonic channel is still appealing because it yields much more signal events and enables reconstruction of the $W$ momenta and thus important kinematics such as the $WW$ invariant mass. Central jet veto and tagging jets requirements have been proposed to reject major QCD backgrounds such as from $t \bar{t}$+jets, and boost the longitudinal fraction of the $W$'s \cite{subtraction}. A persistent background is $W$+jets where the $W$ decays leptonically. In a signal event, the hadronically decaying $W$ is highly boosted at high energies, thus behaves as a single jet in a collider detector, which we call a $W$ jet. In order to reject the $W$+jets background, it is essential to distinguish a $W$ jet from a QCD jet initiated from a quark or a gluon, which has been a focus of the recent jet substructure studies (see \cite{boost2011} for a review). Nevertheless, the latest work before ours on $WW$ scattering using jet substructure and including all major backgrounds\footnote{See Ref.~\cite{Jager:2013mu} for a recent study of the signal alone.} dates back more than a decade ago \cite{Butterworth:2002tt}, where a prototype of the filtering/mass drop method \cite{filtering} was applied to reject QCD backgrounds. In this work we apply the more advanced multivariate $W$ jet tagging method proposed in an earlier paper \cite{wtag}. We find the resultant sensitivity to the beyond-the-SM signal in $W_LW_L$ scattering is better than those in the literature, including those based on the dileptonic channel \cite{Ballestrero:2010vp, Doroba:2012pd}. Our results also suggest that, at the high luminosity 14 TeV run of the LHC, $W_LW_L$ scattering has a good sensitivity to the anomalous $hWW$ coupling.

The rest of the paper is organized as follows. In Section \ref{sec:parton_level} we give definitions of the signal and the backgrounds, and demonstrate at parton level how to suppress the electroweak backgrounds which includes the  $W_LW_T$ and $W_TW_T$ components of $WW$ scattering. In Section \ref{sec:jet_level} we present results from a jet level analysis using the $W$ jet tagging method, taking into account all major backgrounds to the semileptonic $W_LW_L$ scattering. Finally we conclude in Section \ref{sec:conclusion}.

\section{$WW$ scattering at parton level}
\label{sec:parton_level}
\subsection{Signal and Backgrounds Definitions}  
At a hadron collider such as the LHC, a $WW$ scattering event is characterized by a $W$ pair produced with a pair of forward and backward tagging jets. Following the ``subtraction'' approach proposed in Refs.~\cite{subtraction}, we define the new physics signal as the event excess in $WW$ scattering over the SM prediction with a light Higgs ($m_h\approx 125\gev$):
\beq
S=\sigma(pp\rightarrow jjWW)_{\text{new physics}}-\sigma(pp\rightarrow jjWW)_{\text{SM}}\label{eq:sigdef}.
\eeq

As mentioned in the introduction, although a 125 GeV Higgs-like particle has been discovered at the LHC, a sizable excess in $WW$ scattering can still emerge if the new particle has non-SM couplings to $W$ and $Z$, and cannot fully unitarize $W_LW_L$ scattering. This happens in a generic class of composite Higgs models, namely, the Strongly Interacting Light Higgs (SILH) \cite{Giudice:2007fh} models. We use a linear realization of the electroweak symmetry breaking, then the low energy effective theory of an SILH model can be written as the SM Lagrangian with a Higgs doublet, plus higher dimensional operators. In particular, the effective Lagrangian of these models includes the following dimension-6 operator as one of the leading terms, which affects $W_LW_L$ scattering at $\sqrt{s_{WW}}\gg m_h$, where $\sqrt{s_{WW}}$ is the invariant mass of the W pair:
\beq
\mathcal{L}\supset\frac{c_H}{2f^2}\partial^\mu(H^\dag H)\partial_\mu(H^\dag H)\label{lagrangian},
\eeq
where $f$ is the characteristic scale of the new physics. This operator results in a modified Higgs-gauge coupling: $g_{\rm eff}=g_{SM}/\sqrt{1+c_H\xi} \approx(1-c_H\xi/2)g_{SM}$, where the approximation holds for small $c_H\xi$, $\xi=v^2/f^2$ ($v=246$ GeV is the Higgs vev) and $c_H\sim O(1-4\pi)$ depending on the underlying theory. As a result, the presence of the operator in Eq.~(\ref{lagrangian}) prevents the exact cancellation of the $E^2$ growth of the $W_LW_L\rightarrow W_LW_L$ scattering amplitude at $\sqrt{s_{WW}}\gg m_h$. Using Goldstone boson equivalence theorem, the cross section of  $W_LW_L\rightarrow W_LW_L$ at $\sqrt{s_{WW}}\gg m_h$ (yet well below a cutoff scale $E_c$ as we will define in Eq.~(\ref{Ec_def})) is found to relate to the Higgsless model ($g_{\rm eff}=0$) by \cite{Giudice:2007fh}:
\beq
     \sigma_{\rm SILH}(W_LW_L\rightarrow W_LW_L) \approx (c_H\xi)^2\sigma_{\rm Higgsless}(W_LW_L\rightarrow W_L W_L),\label{eq:xsec_relation}
\eeq
where the Higgsless model cross section can be obtained by fixing $v=246\gev$ and taking the Higgs mass to infinity.
Given the observation that the newly discovered particle has Higgs-like couplings to the gauge bosons, a pure Higgsless assumption may not be realistic. However, it is convenient to take the Higgsless limit as one of our benchmark points to evaluate the performance of our method. 

Before moving on, we comment on the validity of the scaling relation given in Eq.~(\ref{eq:xsec_relation}). If the scaling continued to high energies, perturbative unitarity would break down at a scale $E_c$ \cite{Han:2009em}:
 \begin{equation}
E_c^2\approx{16\pi v^2}/{c_H\xi}.\label{Ec_def}
\end{equation}
 For example, $E_c \approx 2$ TeV for $c_H\xi=0.6$. In the Higgsless limit, this happens at a lower $E_c \approx$ 1.2 TeV. Around these scales, contributions from higher dimensional operators (and potentially heavy resonances) need to be included to preserve unitarity. For the case with a heavy Higgs, the effect of these higher dimensional operators have been systematically studied in \cite{Espriu:1994rm,Herrero:1993nc} with the Effective Chiral Lagrangian approach. For the SILH model where Higgs is a light composite scalar, the effective Lagrangian is given in Refs.~\cite{Giudice:2007fh, effective}, while the coefficients of the effective operators depend on the underlying dynamics. Since the focus of this work is to demonstrate the impact/improvement on $WW$ scattering studies using the $W$-jet tagging technique, for simplicity, we will not include higher order operators and adopt the following presciption to take into account the unitarity bounds: for our final results, we cut off all signal events with $\sqrt{s_{WW}}\gtrsim E_c$. The simple parametrization given in Eq.~(\ref{eq:xsec_relation}) should of course not be seen as an accurate result for the SILH models, in particular at $\sqrt{s_{WW}}\gtrsim E_c$. Nonetheless, this approximation may be the easiest yet a sensible way for a phenomenological/experimental study as is our focus in this work. This simplified approach was also used in the existing literature such as Ref.~\cite{Han:2009em}. In general this may result in a conservative estimate for the signal significance, compared with the case when higher dimensional operators or a resonance(s) is included. Nonetheless we will also quote the results when a unitarity limit is not imposed, which allows us to compare to previous studies under the same assumption (without the $E_c$ cut), such as in Refs.~\cite{Ballestrero:2008gf, Ballestrero:2010vp, Doroba:2012pd}. As we will show, due to parton distribution function (PDF) suppressions, the signal diminishes very quickly when $\sqrt{s_{WW}}\gsim 2$ TeV even for the case without a unitarity cutoff.  Therefore, for a smaller $c_H\xi$, thus a larger $E_c$, the two approaches (with or without a unitarity cutoff) yield similar results, and our results are less sensitive to contributions from higher dimensional operators. Given that the current data constrain the anomalous $hWW$ coupling at $1 \sigma$ level to be within $\sim15\%$ (corresponding to  $c_H\xi\approx0.3$) of the SM value \cite{Chang:2013aya}, this may indeed be the case. There is also a lower limit on $\sqrt{s_{WW}}$ for Eq.~(\ref{eq:xsec_relation}) to be valid, namely, Eq.~(\ref{eq:xsec_relation}) is a good approximation only when $\sqrt{s_{WW}}\gg m_H/c_H\xi$ \cite{Contino:2010rs}. (This lower limit is lower than the upper limit given in Eq.~(\ref{Ec_def}), as long as $c_H\xi\gtrsim 0.01$.) As we will see, the LHC can only probe significant deviations ($c_H\xi\sim 0.3$), and a $\sqrt{s_{WW}} >850\gev$ in our analysis helps to ensure the lower limit is satisfied. On the other hand, for smaller $c_H\xi\lesssim O(0.1)$ that may be explored at a high luminosity/energy LHC, one may need to choose a higher $\sqrt{s_{WW}}$ cut.

As demonstrated in Ref.~\cite{Doroba:2012pd}, unlike $W_LW_L$ scattering, the cross sections of $W_TW_T$ and $W_LW_T$ productions are largely independent of the Higgs mass, and almost the same in the Higgsless model and the SM. Moreover, the cross sections for polarization flipping processes are negligibly small (about 3 orders of magnitude smaller than polarization conserving ones \cite{Doroba:2012pd}). Therefore, to a good approximation, except for $W_LW_L\rightarrow W_LW_L$, other contributions to the signal as defined in Eq.~(\ref{eq:sigdef}), are cancelled out. Our signal definition for the SILH models can then be rewritten as
\begin{eqnarray}
S_{\rm SILH}&=&\sigma(pp\rightarrow jjW_LW_L)_{\rm SILH}-\sigma(pp\rightarrow jjW_LW_L)_{\rm SM}\nonumber\\
 &\approx& \sigma(pp\rightarrow jjW_LW_L)_{\rm SILH}.\label{eq:sigdef_SILH}
\end{eqnarray}
In the second line of Eq.~(\ref{eq:sigdef_SILH}), we have ignored the SM contribution to $W_LW_L$ production, which is valid only when we require forward and backward tagging jets to select events produced from $WW$ scattering, and large $WW$ invariant mass to ensure $\sigma_{\rm SILH}(W_LW_L\rightarrow W_LW_L)\gg\sigma_{SM}(W_LW_L\rightarrow W_LW_L)$. 
In this case, the SILH signal is related to the Higgsless signal by a simple rescaling (using Eq.~(\ref{eq:xsec_relation})):
\beq
  S_{\rm SILH}{(c_H\xi)}\approx(c_H\xi)^2S_{\rm Higgsless}\label{eq:signalscaling}.
\eeq
The signal definition and the scaling in $c_H\xi$ are illustrated in Fig.~\ref{fig:mww}.
\begin{figure}[htbp]
\begin{center}
\begin{tabular}{cc}
\includegraphics[width=0.5\textwidth]{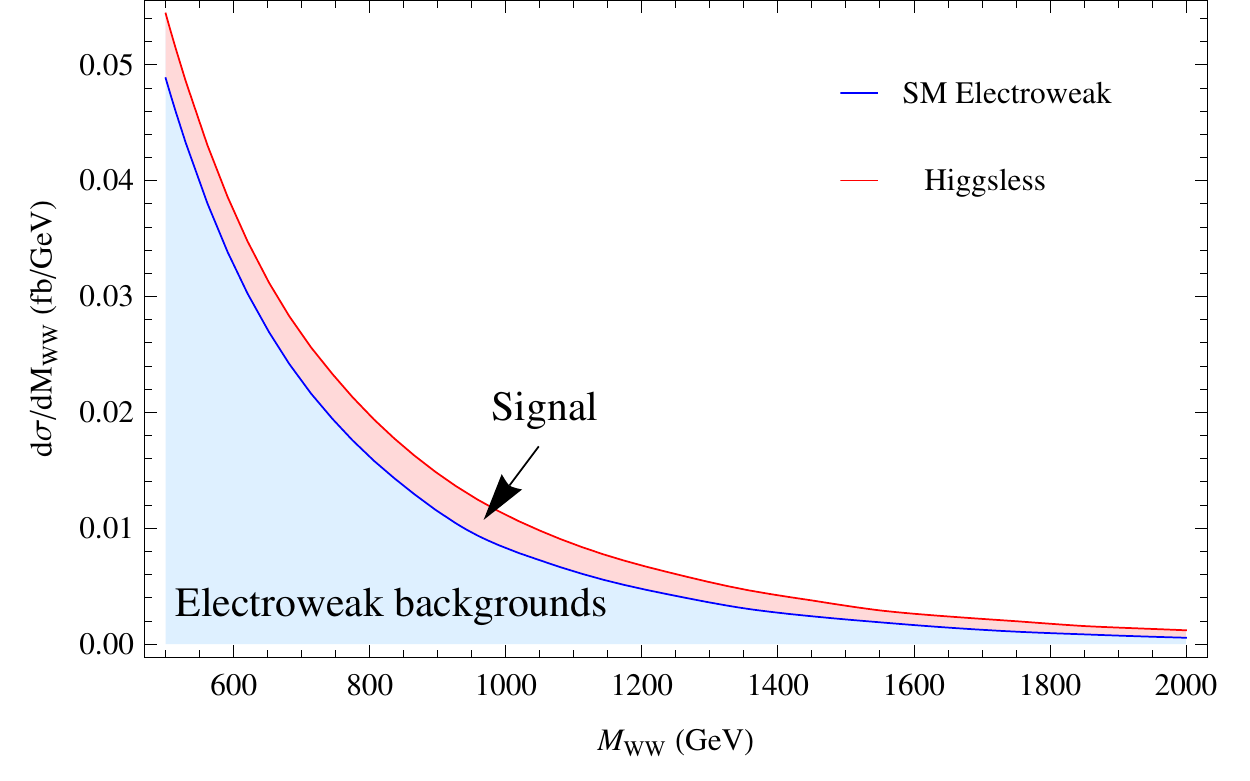}
& \includegraphics[width=0.5\textwidth]{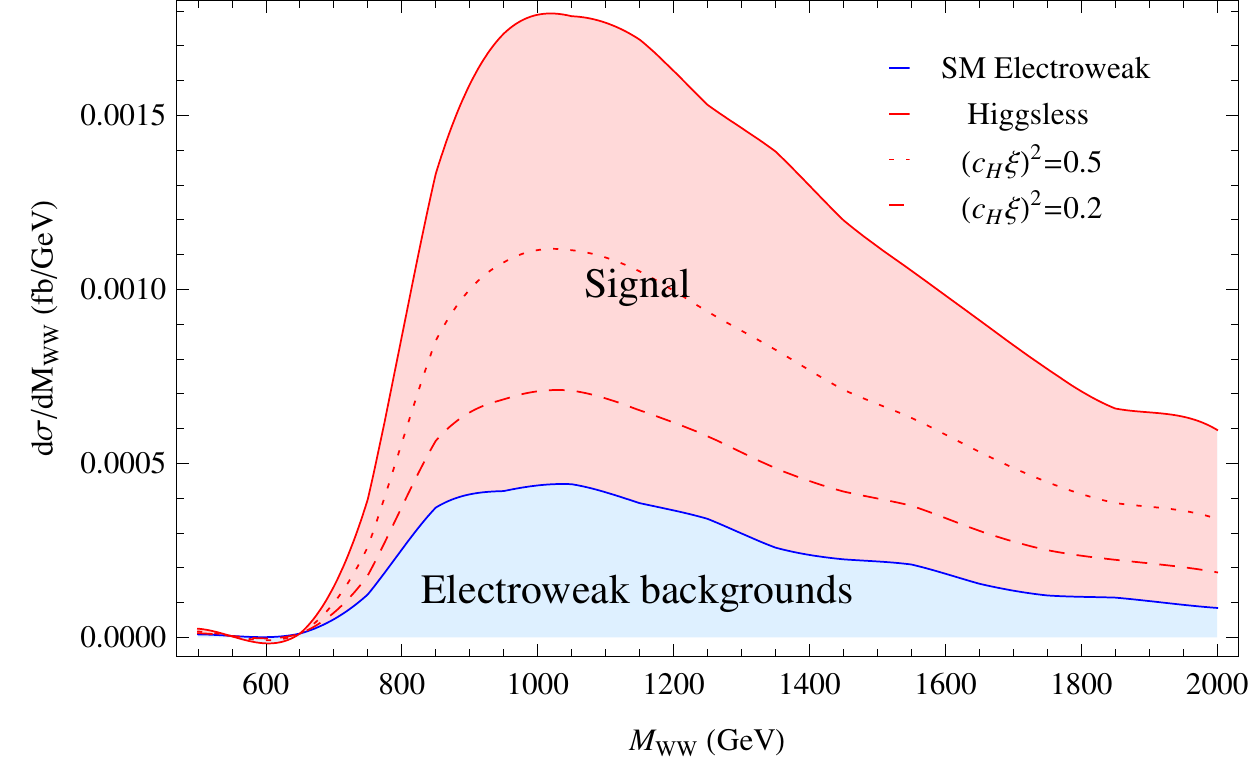}
\end{tabular}
\caption{$WW$jj signals as a function of $m_{WW}$: before (left) and after (right) a $p_T^W>350\gev$ cut on both $W$'s. Only electroweak processes are included in the plots. The blue solid curve denotes the $WW$jj cross section in the SM with a light Higgs ($M_H=125\gev$). The red solid curve is for the Higgsless model, while the dotted and dashed lines are for SILH with $(c_H\xi)^2=0.5$, $(c_H\xi)^2=0.2$, respectively. The signal (light red area) is defined as the difference between the Higgsless/SILH and the SM. Tagging jets are required (Eq. (\ref{eq:generation_cuts})).}
\label{fig:mww}
\end{center}
\end{figure}

In light of this simple scaling relation we will focus on analyzing the LHC sensitivity for the Higgsless assumption in the following discussions. Then we use Eq.~(\ref{eq:signalscaling}) to estimate the sensitivity of the SILH case parametrized by $c_H\xi$. Strictly speaking, in the SILH models Eq.~(\ref{eq:signalscaling}) only holds when $c_H\xi\ll 1$ (but not too small to make the SM $W_LW_L\rightarrow W_LW_L$ non-negligible). Here, we simply take $(c_H\xi)^2$ as a measure of the signal strength and use the scaling relation in Eq.~(\ref{eq:signalscaling}) even for larger  $(c_H\xi)^2$ values.

Now we move on to consider relevant backgrounds. Since we are interested in the excess over the SM, the SM $WW$ scattering events become an irreducible background. These events come from the diagrams involving only electroweak couplings\footnote{Strictly speaking, these diagrams cannot be completely separated from those involving QCD couplings with the same final state particles due to interferences. However, the interferences are tiny in the phase space we are interested in \cite{Doroba:2012pd}.},  which are present regardless whether the $W$'s decay hadronically or leptonically. We will call these backgrounds EW backgrounds. Moreover, as mentioned, in this article we focus on exploring the $WW$ scattering signal in the semileptonic decay channel. Therefore inevitably, we also need to consider processes involving QCD couplings, dominated by $t\bar t$+jets and $W$+jets.  As we will show in the next section, by using our multivariate $W$ jet tagging method, we are able to reduce the QCD backgrounds to a similar level as the EW backgrounds and the final signal sensitivity becomes better than the dileptonic channel. In the next subsection, we will first consider how to reduce the EW backgrounds by examining their differences from the signal at the parton level. In Section \ref{sec:jet_level}, we will include all backgrounds at the jet level and obtain our final results.  

\subsection{Suppressing the electroweak backgrounds -- parton level}
 We generate $pp\rightarrow WWjj\rightarrow l\nu jjjj$ events with MadGraph 5 \cite{madgraph5} at the parton level, for the SM with a 125 GeV Higgs, and for the Higgsless model by turning off the SM diagrams containing the Higgs boson. All charge combinations for the $W$ pair are included. In this section, we only consider backgrounds with pure electroweak couplings, which are present regardless how $W$'s decay, while postponing the discussions on QCD backgrounds to the next section. For illustration, we discuss the Monte Carlo truth at the parton level, which can be straightforwardly applied to the jet level analysis. 

At the generation level, the following  cuts are applied on the two tagging jets to improve the efficiency:

\beq
|\eta_j|<5, \ \ p_T^j>10\gev, \ \ E_j>150\gev, \ \ M_{jj}>300\gev.\label{eq:generation_cuts}
\eeq

At the analysis stage we further employ the following selection criteria to enhance the signal-background ratio.
\bit
\item Two tagging jets $j_1, j_2$ with
\beq
 2<|\eta|<5,  \ \ p_T>25\gev, \ \ E>340\gev \ \ \rm{ and } \ \ \eta_{j_1}\cdot\eta_{j_2}<0. \label{eq:tagging_jets}
\eeq
\item $p_{T}^{W}>350\gev$ for both $W$'s. 
\item The two partons from the $W$ decay have a $p_T$ ratio $> 0.1$ (lower/higher), for both $W$'s.
\item $m_{WW}>850\gev$.
\eit
We summarize the signal and background cross sections after each cut in Table \ref{tab:partonresult}. Note that we have chosen the cuts to be similar to the jet level cuts we use in Section \ref{sec:jet_level}, although other values of the cuts may also suffice for the illustration here. 

The tagging jet cuts are used to select the events from $WW$ scattering. After the tagging jet cuts, we expect the signal-background ratio to increase at higher $WW$ invariant mass due to the $\sim s_{WW}^2$ growth in the  signal $W_LW_L\rightarrow W_LW_L$ cross section. However, from Fig.~\ref{fig:mww} we also see even at 1 TeV, the remaining events are still dominated by the SM backgrounds, which are mainly from $W_TW_T$ and $W_TW_L$ productions. For example, if we apply an $M_{WW}>850\gev$ cut directly after the tagging jet cuts,  we get a signal-background ratio of 0.49. 

To further reduce the background, we utilize the observation \cite{Doroba:2012pd} that the angular distribution for the $W_LW_L\rightarrow W_LW_L$ scattering in the Higgsless model is different from $WW\rightarrow WW$ (including all polarizations) in the SM, namely, with respect to the incoming $W$ direction, the scattered $W$'s are more central in the former case than in the latter. This effect is also more significant at higher $\sqrt{s_{WW}}$ \cite{Doroba:2012pd}. Since the incoming $W$'s are radiated close to the beam line, this results in a larger $p_T^W$ for the signal than for the background in the lab frame. The $p_T^W$ distributions after the tagging jet cuts are shown in Fig.~\ref{fig:ptw}. We then choose a $p_T^W>350\gev$ cut, and obtain the right-side plot in Fig.~\ref{fig:mww}. The signal-background ratio becomes 3.0. For comparison, the corresponding plot before the $p_T^W$ cut is shown on the left side of Fig.~\ref{fig:mww}.  After the $p_T^W$ cut, the $W_LW_L$ component is found to be about 4\% of the 0.54 fb SM EW backgrounds, which is negligibly small compared with the signal. This justifies the second line in Eq.~(\ref{eq:sigdef_SILH}).  

\begin{table}
\begin{tabular}{|c|c|c|c|c|c|}
\hline
&Initial $\sigma$ (fb)&tagging jets& $p_T^W>350\gev$&$p_T$ ratio $>0.1$ &$m_{WW} >850\gev$\\
\hline
\centering Higgsless &252 & 54.9 & 2.15 & 1.84 & 1.74\\
\hline
\centering SM EW& 236 & 48.8 & 0.54 & 0.37 & 0.34\\
\hline
\centering Signal ($\sigma_{\rm Higgsless} - \sigma_{SM}$)&16.0 & 6.07 &1.61 & 1.47 &1.40\\
\hline\hline
\centering$S/B$&0.068 & 0.124 & 2.97 & 3.97 & 4.13
\\\hline
\end{tabular}
\caption{Parton level cross sections of the signal and the SM electroweak backgrounds (in fb) in the semileptonic channel, and the signal-background ratio after each cut. }\label{tab:partonresult}
\end{table}

\begin{figure}[htbp]
\begin{center}
\begin{tabular}{cc}
\includegraphics[width=0.5\textwidth]{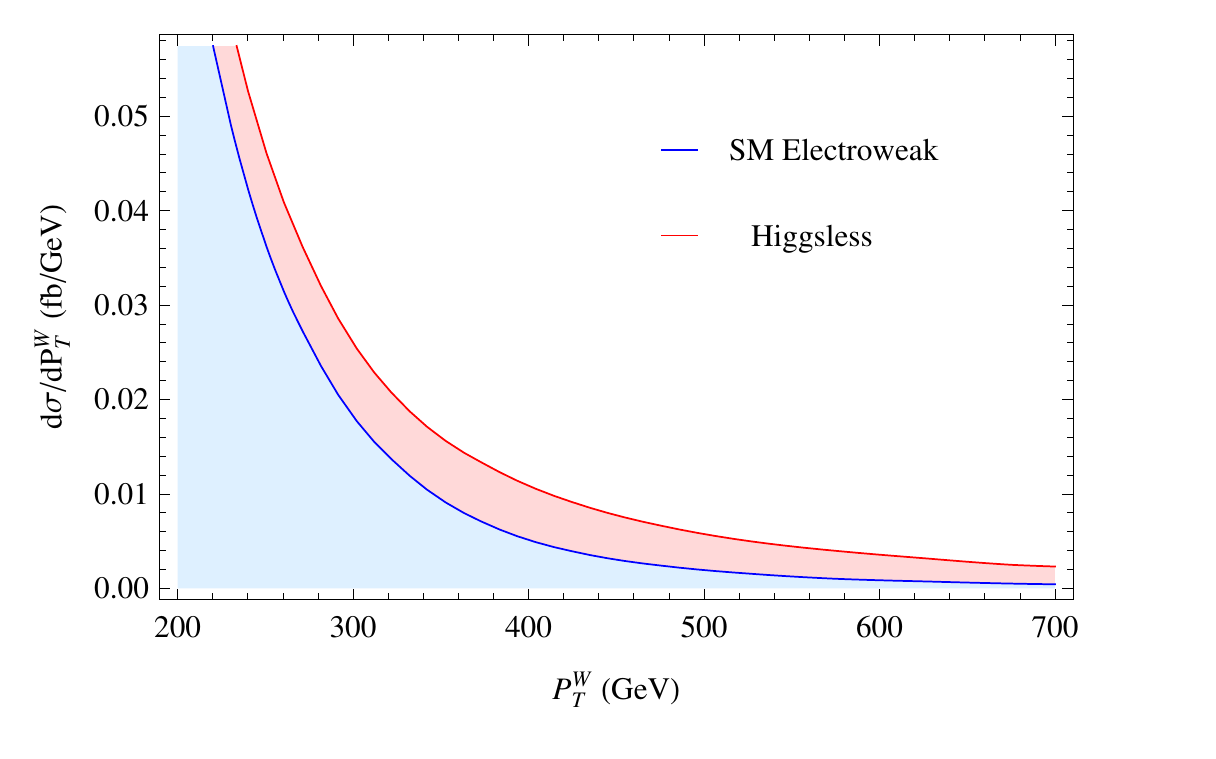} &
\includegraphics[width=0.45\textwidth]{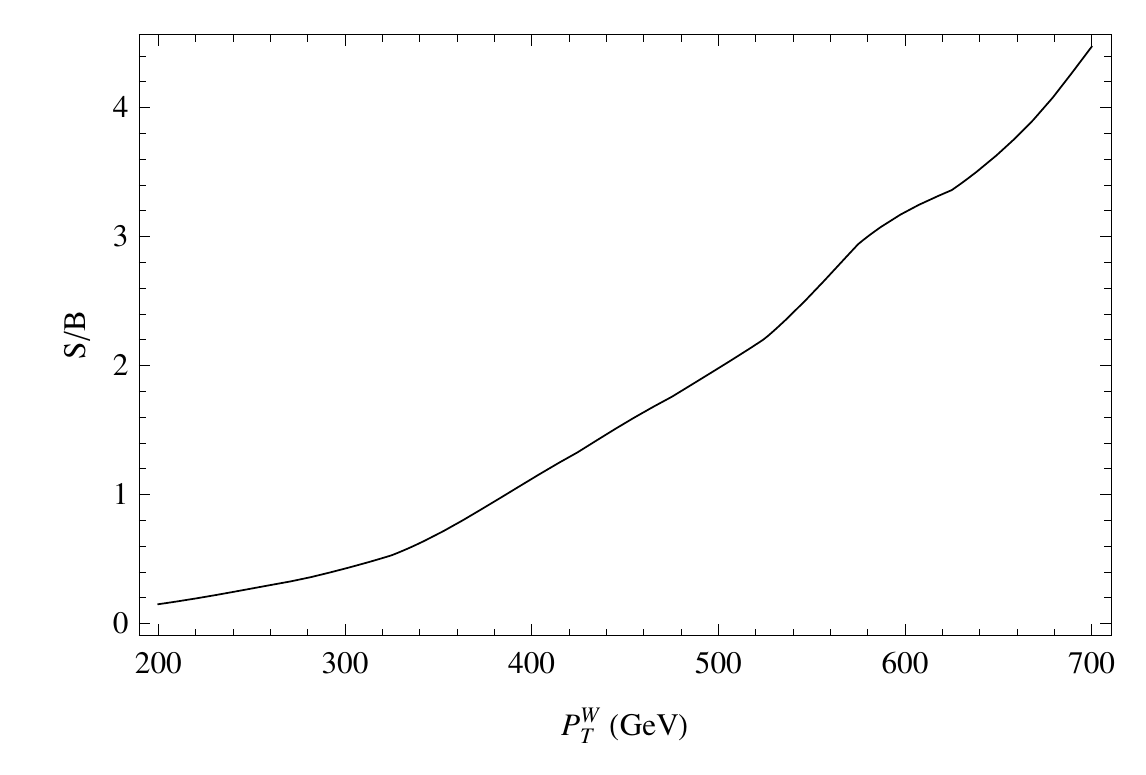}
\end{tabular}
\caption{Left: cross sections for the Higgsless model and for the SM electroweak backgrounds after tagging jet cuts, as a function of $p_T^W$; right: the signal-background ratio, $(\rm{d}\sigma_{Higgsless}-\rm{d}\sigma_{SM})/\rm{d}\sigma_{SM}$, as a function of  $p_T^W$.}
\label{fig:ptw}

\end{center}
\end{figure}

\begin{figure}[htbp]
\begin{center}
\includegraphics[width=0.6\textwidth]{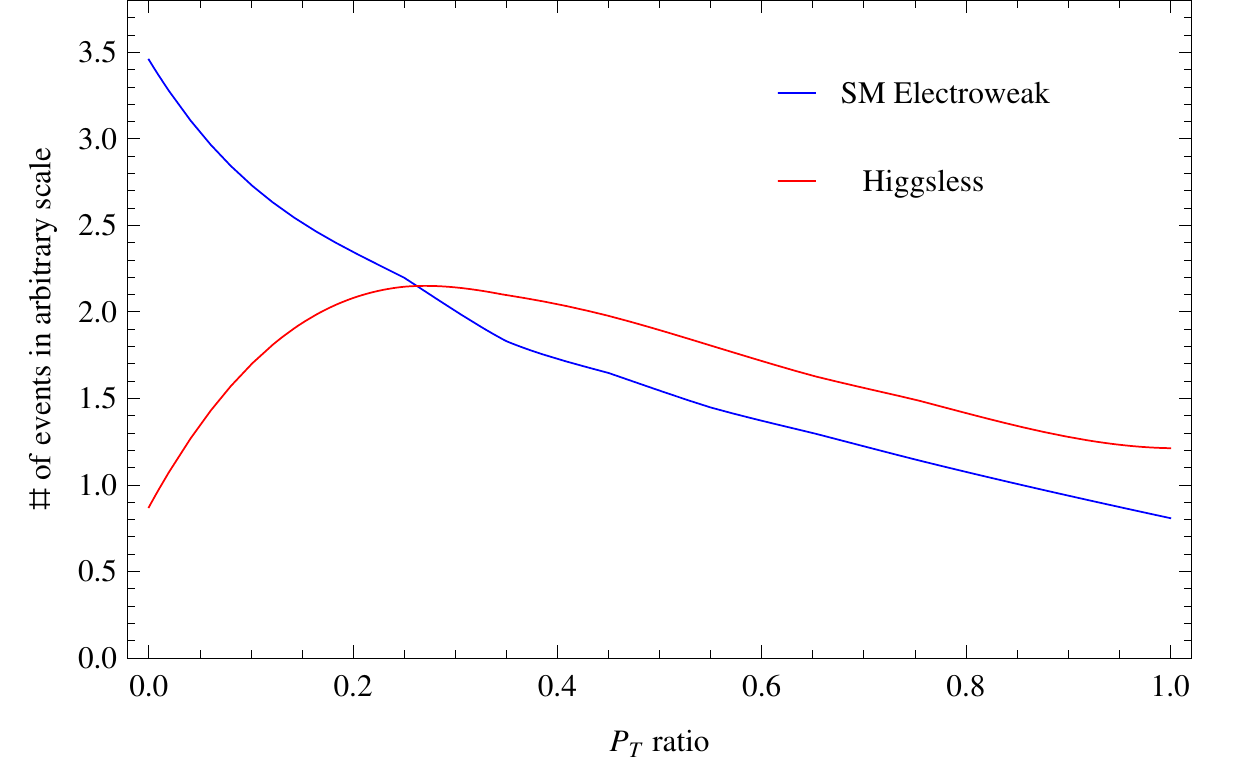}
\caption{The ratio of the lower $p_T$ to the higher $p_T$ for the two partons from $W$ decay.}
\label{fig:ptratio}
\end{center}
\end{figure}

 Our final control over the EW backgrounds comes from the $W$ polarizations: the signal only contains longitudinal $W$'s, while the background contains both longitudinal and transverse $W$'s. This results in different angular distributions for the $W$ decay products. For a longitudinal $W$, the decay products tend to move perpendicularly to the $W$ moving direction, in the $W$ rest frame; for a transverse $W$, one parton tends to move along the $W$ moving direction while the other one against it. Therefore, when the $W$ is boosted, the momenta of the two decay products tend to be more balanced for a longitudinal $W$ than for a transverse $W$. We see this effect by examining the $p_T$ ratio of the two decay products, as shown in Fig.~\ref{fig:ptratio}. The Higgsless model, dominated by longitudinal $W$'s, has fewer events in the low $p_T$-ratio region. Therefore, we can put a cut on $p_T$-ratio to further increase $S/B$, as shown in Table \ref{tab:partonresult}. In practice, when a highly boosted $W$ decays hadronically, the two partons yield two subjets. We then require a more balanced momentum configuration for the two subjets. Coincidentally, this requirement is also essential for reducing QCD jet backgrounds since QCD jets typically do not contain balanced subjets.

\section{$WW$ scattering with jet substructure}
\label{sec:jet_level}

The dileptonic decay channel for a $W$ pair is considered ``gold-plated'' conventionally, since it has smaller backgrounds than the semileptonic channel or the all hadronic channel. However, it suffers from a smaller branching ratio and thus worse statistics. In addition, the presence of two missing neutrinos makes it hard to reconstruct the kinematics. In this article, we will concentrate on the semileptonic channel, and show that it has a better discovery potential than the dileptonic channel. 

Due to the fact that we are studying $WW$ scattering at high $\sqrt{s_{WW}}$ around 1 TeV, the hadronic $W$ is highly boosted and thus it is important to distinguish it from a QCD jet. We will utilize the state-of-art $W$ tagging method as described in Ref.~\cite{wtag}.
Here it is helpful to first briefly review the physics behind $W$-jet tagging and the major results in Ref.~\cite{wtag}. There are two major differences between a $W$-jet and a QCD jet. Firstly, a $W$ jet contains two hard subjets ({\it i.e.}, subregions where the jet energy is concentrated), originated from the two quarks in the $W$ decay, while a QCD jet usually has only one hard subjet. Various jet algorithms have been proposed for identifying the hard subjets, including filtering \cite{filtering}, pruning \cite{pruning} and trimming \cite{trimming}. Secondly, the $W$ boson is a color singlet particle, consequently all QCD radiation from the decay of a boosted $W$ is confined in a small cone around the $W$ momentum direction.  On the other hand, a QCD jet is initiated from a color triplet or octet, which is color-connected to the beam or the other side of the event. Therefore, the radiation of a QCD jet is usually much more diffuse. This difference is manifest, for example, by examining how the jet mass (or $p_T$) grows when the jet radius increases from the jet energy center \cite{wtag}. This is well captured by the $R$-core variables defined in Ref.~\cite{wtag}. The most powerful discriminant is built upon variables sensitive to the above two major differences using a multivariate analysis (MVA), as shown in Ref.~\cite{wtag}, where a factor of $\sim5$ is achieved in the improvement of statistical significance ($S/\sqrt{B}$) for $p_T\ge200\gev$. We give more information of the multivariate W-tagging method in the Appendix, while refer readers to Ref.~\cite{wtag} for further details.

\subsection{Event simulation and selection}
As described in Sec.~\ref{sec:parton_level}, electroweak processes involving $WW$ + 2 jets are simulated with Madgraph 5 at the parton level for the Higgsless model and the SM with a light Higgs. Minimal cuts in Eq.~(\ref{eq:generation_cuts}) are applied when generating the events.  We then use Pythia 8 \cite{pythia8} to add parton showers and hadronization. Besides the electroweak backgrounds discussed in Sec.~\ref{sec:parton_level},  the process also receives backgrounds from $t\bar t$ + jets and $W$ + jets, in which QCD couplings are involved. For simplicity we will slightly abuse the nomenclature and call these QCD backgrounds although they also involve electroweak couplings. To estimate these backgrounds, we simulate the $2\rightarrow 2$ processes of $W$ + 1 jet and  $t\bar t$ with Pythia 8 and have the extra jets generated from parton showers. In addition, in Sec.~\ref{sec:parton_level} we only included pure electroweak processes for $WW$ + 2 jets, while omitting processes with QCD couplings. As discussed in Ref.~\cite{Doroba:2012pd}, the latter is a minor background once we require two forward/backward tagging jets. For completeness, in this section we include this latter background as part of our QCD backgrounds and treat it similar to $t\bar t $ + jets and $W$+jets, namely, we generate $pp\rightarrow WW$ events and add parton showers with Pythia 8.  As mentioned in Ref.~\cite{Iordanidis:1997vs}, the tagging jets in the QCD backgrounds may not be modeled precisely from the parton shower approach. A more accurate background estimate may be obtained using other tools such as Alpgen \cite{alpgen} or MC@NLO \cite{mcnlo}. Because the main purpose of this paper is to show the relative improvement in the study of $WW$ scattering by using $W$ jet tagging, we stick to the Pythia 8 simulations. 

At the generation level, a $p_T>150\gev$ cut is applied for the $W$ + 1 jet and $WW$ processes and $p_T>100\gev$ for $t\bar t$. We assume a 50\% b-jet tagging efficiency and veto events that contain one or more b-jets. This cuts off 75\% $t\bar t$ events. We use the default settings of Pythia 8, in which ISR, FSR and the underlying event are included, while pileup is not included\footnote{Pileup will affect the performance of our $W$ jet tagging method. However, it is shown in Ref.~\cite{pileup} that the effect can be largely corrected using a subtraction scheme. As an example, Ref.~\cite{pileup} shows the performance of top tagging degrades by about 10\% in terms of $S/\sqrt{B}$ when the number of pileup vertices is 60. It remains to be seen whether this type of subtraction performs similarly well when incorporated in $W$ jet tagging. Moreover, we note that some of the pileup effect can be avoided by using variables based on the tracking information \cite{tracking}, and requiring the tracks to originate from the primary vertex.}. After hadronization, all visible stable particles are grouped in $0.1\times 0.1$ bins on the $(\eta, \phi)$ plane, corresponding to hadronic calorimeter resolutions.

We then reconstruct the signal and background events and impose selection cuts with the following procedure.  At this stage, we keep some of the critical cuts unfixed and later will vary them to obtain the optimal results according to the case being studied.
\begin{enumerate}
\item{Lepton and missing momentum from $W\rightarrow\ell\nu$}\\
Isolated leptons are defined similar to Ref.~\cite{atlas-higgs-semi}: an isolated muon or electron has to pass the cuts: $p_T>20\gev$ and $|\eta|<2.4$. The sum of the transverse energies around the lepton in a cone $\Delta R=\sqrt{\Delta \phi^2+\Delta\eta^2} < 0.3$ must satisfy $\sum E_T < 0.14p_T^{\text{lep}}$. The sum of the transverse momenta for all tracks around the lepton in a cone $R<0.3$ must satisfy $\sum_{\text{tracks}} P_T < 0.13p_T^{\text{lep}}$. The energies of all isolated leptons are removed from the corresponding calorimeter cells for jet clustering. The isolated lepton with the highest $p_T$ is assumed to be the lepton from the leptonic $W$ decay, which is required to satisfy $p_T^{\text{lep}}>40\gev$. Furthermore, we apply a cut on the missing transverse momentum: $\slashchar{p}_T>40\gev$, where $\slashchar{p}_T$ is defined as the negative sum of the visible momenta, including those of the isolated leptons and those deposited in the calorimeter cells.

\item{Neutrino momentum reconstruction}\\
 Assuming the transverse momentum of the neutrino from $W\rightarrow \ell\nu$ is given by the missing transverse momentum, we use the $W$ mass shell constraint to solve for the neutrino momentum in the beam direction, $p_z^\nu$. This method gives us a quadratic equation leading to either 0 or 2 real solutions. We discard events without real solutions. In the case of two real solutions, we keep the solution that gives the smaller $|p_z^\nu|$. The reconstructed leptonic $W$ is required to satisfy
\begin{equation}
p_T^{W\rightarrow \ell\nu}> p_{T,\text{cut}}^W.
\end{equation}

\item{$W$ jet tagging}\\
 We use the $Cambridge/Aachen$ algorithm in FastJet \cite{fastjet} and cluster the calorimeter cells to fat jets with $R=1.2$ . The leading jet is taken to be the $W$ jet candidate which is required to pass the filtering/mass drop procedure described in Appendix \ref{app:wtag}. After filtering, the jet mass is required to fall in the window $(60,100)\gev$. We take the jet momentum after filtering as the $W$ momentum, which is required to pass
\begin{equation}
p_T^{W\rightarrow jj, filter} > p_{T,\text{cut}}^W.
\end{equation}
To further discriminate $W$ jets from QCD jets, we use the $W$ jet tagging method described in Ref.~\cite{wtag}. In this method, Boosted Decision Trees (BDT), a multivariate event classifier, is used to discriminate $W$ jets from QCD jets. The signal/background efficiency is variable by varying the BDT cuts. In Ref.~\cite{wtag}, we trained jets in different $p_T$ bins (bin size = 50 GeV) separately, therefore the BDT cuts for different $p_T$ bins can be chosen independently. In this work, to simplify our procedure, we choose the BDT cuts so that the nominal signal efficiency $\varepsilon_S^{\rm BDT}$ is the same for all $p_T$ bins. Here $\varepsilon_S^{\rm BDT}$ varies from 0 to 1, where 1 means we keep all the signal events that have passed the filtering/mass drop procedure. Note that the nominal signal efficiencies in Ref.~\cite{wtag} were obtained from the SM $WW$ pair production, where the $W$'s are dominantly transversely polarized. When the $W$'s are dominantly longitudinal, as in our signal events, the actual efficiency is higher. 

\item{Central jet veto}\\
We remove the calorimeter cells constituting the $W$ jet after filtering. The remaining calorimeter cells of the fat jet, as well as all other calorimeter cells are reclustered with a smaller jet radius, $R=0.4$, using the anti-$k_t$ algorithm.  We veto events with one or more central jets that satisfy $|\eta|< 2$ and $p_T > p_{T,\cut}^{\cj}$.

\item{Forward/backward tagging jet cuts}\\
 We look for a forward tagging jet with $5>\eta>2$ and a backward jet with $-5<\eta <-2$. The energies of the two tagging jets are required to pass the following cuts:
\begin{equation}
E^{\text{forward}} > E_{\cut}^{\fb}, \  \ E^{\text{backward}} > E_{\cut}^{\fb}.
\end{equation}  
\item{$M_{WW}$ cut}\\
Finally, we apply a cut on the $WW$ invariant mass, 
\begin{equation}
M_{WW}>M_{WW,\cut}.
\end{equation}
\end{enumerate}
This cut is useful, especially for reducing the QCD backgrounds, which drop faster than the signal when $M_{WW}$ increases.

\subsection{Results}
In the above procedure, we did not specify the values for the following cuts:  the $p_T$ cuts for the leptonic and hadronic $W$'s, which we take to be equal, $p_{T,\cut}^W$; the BDT efficiency, $\varepsilon_S^{\rm BDT}$; the $p_T$ cut for  the central jet, $p_{T,\cut}^{\cj}$; the energy cuts for the forward and backward tagging jets, which we take to be equal, $E_{\text{cut}}^{\fb}$; and the $WW$ invariant mass cut, $M_{WW,\cut}$. We then vary these cuts to achieve the best statistical significance, $S/\sqrt{B}$, where $B$ includes all the EW and QCD backgrounds. Note that we may choose different $p_T$ cuts for the hadronic $W$ and the leptonic $W$, as well as different cuts for the leading and the next-to-leading forward/backward jet. The BDT efficiency for each $p_T$ bin can also be chosen independently. Allowing more cuts to be variable, we may obtain better results than presented in this article.

For the Higgsless model, the best cuts are found to be:
\begin{equation}
p_{T,\cut}^W= 350\gev, \ \ \varepsilon_S^{\rm BDT}=0.54, \ \  p_{T,\cut}^{\cj}=70\gev, \ \ E_{\text{cut}}^{\fb}=340\gev, M_{WW,\cut}=850\gev.
\end{equation} 
We see that the optimal central jet veto cut is rather mild: $p_{T,\cut}^{\cj}=70\gev$. This cut was mostly designed to reduce the $t\bar t$ background because it contains extra jet activities in the central region. In our approach, this cut is partially redundant to the $W$ tagging method, where a $W$ jet candidate is required to have few extra hadronic activities, as expected for a jet originating from a color singlet particle. 

The resultant $M_{WW}$ distributions (before the final $M_{WW}$ cut) for the signal and the backgrounds are shown in Fig.~\ref{fig:mww_jet}. After the $M_{WW}>850\gev$ cut, the cross section for the Higgsless signal is 0.49 fb and for the total backgrounds 0.24 fb. As discussed in Section \ref{sec:parton_level}, perturbative unitarity breaks down for the Higgsless assumption at $M_{WW} \sim 1.2$ TeV. To estimate the effect, we cut off signal events with $M_{WW} < 1.2$ TeV at the parton level, and corresondingly apply the same cut on the reconstructed $M_{WW}$. After this cut, the cross section of the signal becomes 0.19 fb and that of the total backgrounds becomes 0.14 fb, which means with 300 fb${}^{-1}$ data at the 14 TeV LHC, we can discover a Higgsless model at the 5 $\sigma$ level (Poisson statistics is used for small numbers of signal and background events). The discovery potential for the SILH model is scaled down by a factor of $(c_H\xi)^2$. However, perturbative unitarity is preserved until a higher scale. In particular, the LHC at the high luminosity (LHC-HL) run with $\sim 3000$ $\rm fb^{-1}$ data can lead to the discovery of an SILH model with $c_H\xi\sim0.3$, corresponding to a $\sim15\%$ deviation from the SM $hWW$ coupling. In this case, we have applied a cutoff of 3 TeV which has almost no effect on the signal due to the rapid decrease in PDFs at large x's. 

Note that in the above estimates, we have not included systematic uncertainties, which are important, particularly for small S/B ratios. For example, when $c_H\xi\sim0.3$, $S/B\approx 0.2$, therefore the systematic uncertainties on the backgrounds have to be reduced to the level of a few percent to make a discovery. A dedicated study is needed to understand whether this is achievable. Here, we speculate on methods for eliminating some of the systematic uncertainties. The QCD backgrounds have large theoretical uncertainties in the cross sections, however, most of them can be eliminated by using data-driven methods when large statistics is available. For example, we may use b-rich $t\bar t$ events, in which signal events are rare,  as a control sample to estimate the $t\bar t$ background in the signal sample. To estimate the size of the $W$ + jet background, the control sample can be chosen as events with very similar kinematics, but with the $W$ candidate jet mass just outside the $W$ mass window. A data driven method is perhaps unavailable for the SM EW background, which contains real VBF events. The current theoretical calculations for the cross section of a 1 TeV Higgs in the VBF channel have $\lesssim 10\%$ uncertainties \cite{higgs-xsec}, which needs to be refined to make it directly usable. However, one may also examine the polarization fractions instead of the total cross sections to reduce the systematic uncertainties, following Ref.~\cite{Han:2009em}. Moreover, we can also achieve a larger $S/B$ by slightly sacrificing the significance. For example, by setting $p_{T,\cut}^W= 440\gev$, $\varepsilon_S^{\rm BDT}=0.36$,  $p_{T,\cut}^{\cj}=90\gev$, $E_{\text{cut}}^{\fb}=400\gev$ and $M_{WW,\cut}=1200\gev$, $S/B$ is increased by more than 100\% and $S/\sqrt{B}$ only decreases by 4\%.

If the systematic uncertainties are under control, in terms of the sensitivity to the anomalous $hWW$ coupling, at 1 $\sigma$ the $3000$ $\rm fb^{-1}$ run has a sensitivity to $c_H\xi\sim0.15$, i.e., a $\sim8\%$ deviation from the SM coupling. Taking $c_H\sim1$, this implies a sensitivity for the composite scale $\Lambda\sim4\pi f$ up to $\sim$ 8 TeV. In comparison, fitting based on precision measurements at the Higgs resonance offers a very different path for probing a possible anomalous $hWW$ coupling. There the 1 $\sigma$ error bar is estimated to be $\sim 5\%$ of the SM prediction at the LHC-HL \cite{Klute:2012pu,Peskin:2012we, CMS-HL}, dominated by systematic uncertainties. Though our results are not without systematic uncertainties, they are from completely different sources. In particular, at the Higgs pole, all Higgs production and decay channels are entangled and have to be fit together, while $WW$ scattering is a direct probe to the $hWW$ coupling, insensitive to uncertainties such as invisible/exotic Higgs decays. Therefore, we see that assisted by the jet substructure techniques, $WW$ scattering in the semileptonic channel can be a complement to the Higgs pole measurement of the $hWW$ coupling. 

\begin{table}[t]
\begin{center}
\begin{tabular}{|c|c|c|c|c|c|}
\hline
 &Higgsless& \begin{tabular}{c}SM EW\\$M_H=125\gevc$\end{tabular} &\begin{tabular}{c} $W$+jets\\$p_T>150\gevc$\end{tabular} & \begin{tabular}{c}$t\bar t$\\$p_T>100\gevc$\end{tabular} &\begin{tabular}{c} $WW$(+QCD jets)\\$p_T>150\gevc$\end{tabular} \\
\hline
$\sigma\times \text{BR}$ (fb) & 250 & 235  & 23.4k & 101k & 650\\
\hline
isolated lepton               & 72.7  & 64.8   & 7681 & 43.2k & 311\\
\hline
\begin{tabular}{c}leptonic $W$ reconstruction\\$p_T^{W\rightarrow \ell\nu}>350\gev$\end{tabular}
                              & 9.37 & 6.63  & 192    & 2.30k & 24.4\\
\hline
 $p_T^{W\rightarrow jj,\text{filtering}}>350\gev$           
                              & 3.31 & 1.49  & 23.6   &121   &8.82\\
\hline
$W$-jet tagging, $\varepsilon_S^{\rm BDT}=0.54$        
                              & 2.08 & 0.81  & 5.38   &13.8    &5.43\\

\hline
\begin{tabular}{c}tagging jets found\\$E>340\gev$ \end{tabular}
                              & 0.73 & 0.14  &0.16    & 0.14 & 0.02\\
\hline
 $p_T^\cj < 70\gev$     
                              & 0.63 & 0.12  &0.06    &0.09 & 0.01\\
                             
\hline
$m_{WW}>850\gev$               & 0.59 & 0.11  &0.04   &0.07  &0.01\\
\hline\multicolumn{6}{|c|}{ $S={\rm Higgsless - SM} =0.49\, \rm  fb$, $B=0.24 \,\rm  fb$}   \\ 
\hline
\end{tabular}
 \end{center}
\caption{Step by step cross sections after each cut. The cross section of $t\bar t$ is multiplied by a factor of 0.25 to account for b-jet veto effect.\label{tab:nevts_higgs}}
\end{table}
\subsection{Comparison to other analyses}
Dileptonic $W$ decay channels, especially from same sign $W$'s, are usually considered as ``gold plated'' due to their lower backgrounds compared to the semi-leptonic channel. We compare our results to the most recent results for the dileptonic channels such as in Ref.~\cite{Doroba:2012pd}. The authors of Ref.~\cite{Doroba:2012pd} showed the important difference in the $W$ $p_T$ distribution between the SM and the Higgsless model, although in practice lepton $p_T$ is used as the discriminator because the $W$s' momenta are not reconstructable in the dileptonic channel. By utilizing the difference in the lepton $p_T$ distribution they obtain better results than previous analyses, for example, \cite{Ballestrero:2010vp}. In particular, the best performance comes from the same sign dileptonic channel. When no unitarity limit is applied, they obtain $S/B\approx 17/3$ for 100 fb${}^{-1}$ from their $jjWW$ simulations, excluding $t\bar{t}$ background.  Their estimate for the $t\bar t$ background is about 1 event per 100 fb${}^{-1}$, where the same sign lepton comes from a b meson decay. Therefore, their final best result is $S/B\approx 17/4$ for the same sign dileptonic channel. Note that adding the opposite sign dileptonic channel only slightly changes the statistical significance because the contribution from $t\bar t$ to opposite sign leptons is much larger.

    In comparison, our result for the Higgsless model is $S/B\approx 49/24$ (without unitarity limit), which is $\sim 20\%$ better in $S/\sqrt{B}$,\footnote{When the number of events is small such as at 100 fb${}^{-1}$, the actual improvement in significance is even larger due to the deviation from Gaussian statistics.} and $\sim 200\%$ more signal events are retained. This result is obtained by scanning the cuts to maximize the significance. Although the signal-background ratio is smaller than the dileptonic channel, if needed, one may choose the cuts to increase $S/B$ without sacrificing significantly the significance. As mentioned in the previous subsection, by using different cuts, we obtain $S/B\approx 20/4.3$ for 100 fb${}^{-1}$. This $S/B$ is similar to the dileptonic channel, and $S/\sqrt{B}$ is only slightly smaller than the optimal result.

Other works on semileptonic $WW$ scattering include Ref.~\cite{Butterworth:2002tt}, where the jet substructure method was proposed. A direct comparison to Ref.~\cite{Butterworth:2002tt} is impossible because they focused on models with resonances which can greatly enhance the signal and consequently they simplified the signal definition by including all $WW\rightarrow WW$ events instead of using the subtraction scheme we employed. Nevertheless, we have incorporated the filtering/mass drop method \cite{filtering} as an essential step in our $W$ jet tagging method, which is an improved version of the $y$-splitter method in Ref.~\cite{Butterworth:2002tt}. For comparison, we did a check by turning off the MVA $W$ jet tagging method and using filtering alone to identify the $W$ jets. Then we repeated the optimizing procedure by scanning over the essential cuts, and obtained the best significance $\sim 6.7$ per 100 fb${}^{-1}$ ($S=40$, $B=25$) which is $\sim 16\%$ lower than the outcome using MVA W jet tagging. A more recent parton-level analysis in the semileptonic channel is given in Ref.~\cite{Ballestrero:2008gf}. However, the jet substructure method was not used in that analysis, consequently a much smaller $S/B$ ratio was obtained which led to pessimism for this channel. 

\begin{figure}[th!]
\begin{center}
\includegraphics[width=0.8\textwidth]{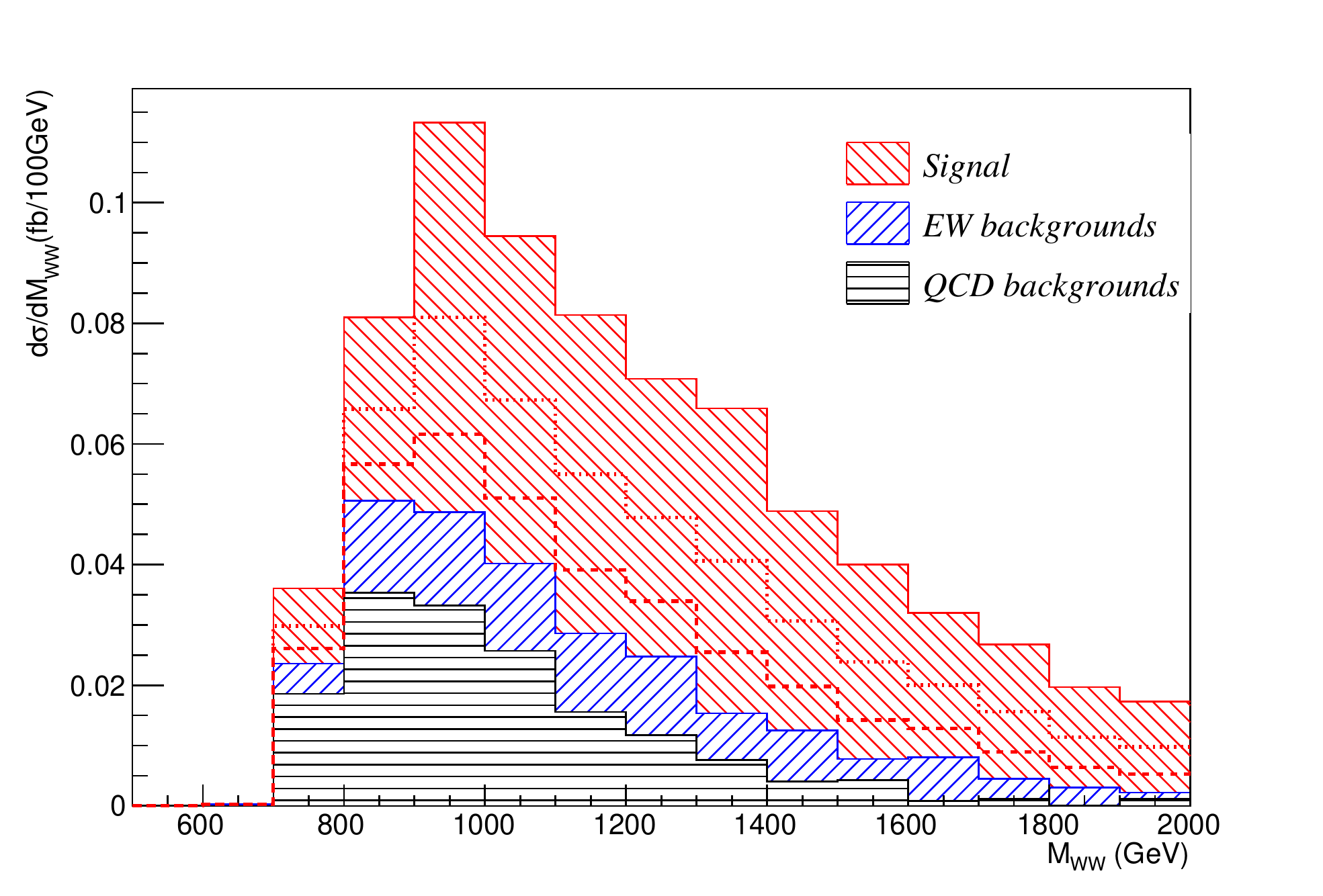} 
\caption{Reconstructed $M_{WW}$ for the signal and the backgrounds. The solid curve indicates the Higgsless signal. The dotted (dashed) curve denotes the SILH signal with $(c_H\xi)^2=0.5$ ($(c_H\xi)^2=0.2$). }
\label{fig:mww_jet}
\end{center}
\end{figure}

\section{Conclusion}
\label{sec:conclusion}

In this article, we have demonstrated that the semileptonic channel can play a major role in extracting new physics from high energy $WW$ scattering at the LHC. In particular, we have considered the strongly interacting light Higgs scenario as a new physics example that induces an excess in $W_LW_L\rightarrow W_LW_L$ scattering, compared to the prediction of the SM with a 125 GeV Higgs. In practice the signal of this model can be approximated based on that of the Higgsless model by a simple rescaling. 

 Compared to the traditionally ``gold-plated'' dileptonic channel, the semileptonic channel is blessed with a larger branching ratio and the possibility to reconstruct important kinematics, although it bears larger backgrounds. Useful schemes such as imposing tagging jets cuts and central jet veto have been found effective for reducing QCD backgrounds, yet still not enough to make the semileptonic channel competitive to the dileptonic channel. We have shown that, assisted by the recently developed jet substructure techniques, in particular the multivariate $W$ jet tagging method \cite{wtag}, we can further greatly reduce the persisting backgrounds from $W$+jets and $t\bar{t}$+jets, while keeping a significant amount of signals. The best statistical significance ($S/\sqrt{B}$) we have achieved is about $20\%$ better than the best existing result for the dileptonic channel. Meanwhile, more than $\sim200\%$ of the signal events are retained, which can benefit further investigations on properties of the new physics. Although the signal-background ratio is moderately smaller than the dileptonic channel, a similar ratio can be obtained by slightly sacrificing the significance (yet still larger than that of the dileptonic channel). These results make the semileptonic channel as good as, or even more promising than the dileptonic channel. Applying the method to the Higgsless model (with a cutoff on $WW$ invariant mass of 1.2 TeV to account for the unitarity bound), we found that the signal can be discovered at the $5\sigma$ level using about 300 fb${}^{-1}$ data at the 14 TeV LHC. For an SILH model with an anomalous Higgs-$W$ coupling, 3000 fb${}^{-1}$ is needed to reach a 5 $\sigma$ discovery (1 $\sigma$ sensitivity) if the deviation from the SM coupling is 15\% (8\%). Therefore, $WW$ scattering is a good complement to the precision measurements for probing $hWW$ coupling, while having less sensitivity to uncertainties such as from invisible/exotic Higgs decays. In addition, our method is also expected to boost the discovery opportunity for heavy resonances responsible for unitarizing $WW$ scattering in the SILH models.
 
 Meanwhile we would like to comment that the analysis in this article is not comprehensive and can be further elaborated by including subleading effects and uncertainties, such as: other channels, e.g. $WZ$ and $ZZ$ scattering; the cross sections at next-to-leading order; the effect of pileup; experimental uncertainties beyond simple geometrical acceptance cuts and hadronic calorimeter resolution, etc. Nevertheless, we do not expect these to change our conclusion, {\it  i.e.}, with the jet substructure techniques, the semileptonic channel is among the best for $WW$ scattering studies and more efforts, both theoretical and experimental, should be devoted to this channel. We hope our work could shed new light on finding new physics from $WW$ scattering: an important yet conventionally challenging process.

\section*{Acknowledgement}
We thank Matthew Schwartz for comments on the manuscript, Roberto Franceschini and Brock Tweedie for useful conversations. Y.~Cui was supported in part by NSF grant PHY-0968854 and by the Maryland Center for Fundamental Physics. Z.~Han was supported in part by DoE grant No.\ DE-FG-02-96ER40969. 

\appendix
\section{$W$ jet tagging}
\label{app:wtag}

As mentioned in the introduction, two major differences make a $W$ jet distinguishable from a QCD jet. Firstly, a $W$ jet contains two sub-regions where the jet energy is concentrated, originating from the two quarks from the $W$ decay. These sub-regions are usually called subjets. Subjets can be conveniently identified by using a recombination jet clustering algorithm: one first finds the constituents of the original fat jet which is obtained with a jet radius $R_\fat$, then uses a smaller $R<R_\fat$ to recluster the constituents, yielding more than one subjets. Usually, the two leading subjets in a $W$ jet correspond to the two partons from the $W$ decay, and have a balanced momentum configuration. On the other hand, QCD splitting tends to produce subjets with  hierarchical momenta. Therefore, a QCD jet usually contains one and only one hard subjet, together with a set of soft subjets. In jet grooming algorithms \cite{filtering, pruning, trimming}, one takes a step further by discarding the soft subjets and keeping only a few hard subjets in the ``groomed jet". After grooming, the mass of the $W$ jet does not change significantly, which is still around the $W$ mass, while that of a QCD jet is often shifted to a small value. One can then use a $W$ mass window cut to eliminate most of the QCD jets.

Due to its effectiveness, the first step in our $W$ tagging method described in Ref.~\cite{wtag} is to use the filtering/mass drop algorithm \cite{filtering} to identify $W$ jet candidates. In particular, we start from fat jets clustered with $R=1.2$ using the Cambridge/Aachen jet algorithm. We then follow the mass drop algorithm in Ref.~\cite{filtering} to find a smaller subjet radius, $R_{\rm filt}$, and use it to recluster the fat jet into subjets. We keep the leading three subjets as the new, filtered jet.  The jet mass after filtering is required to be within the mass window of $(60, 100)\gev$. The filtering/mass drop parameters are optimized to maximize the significance, $S/\sqrt{B}$, where $S$ is the number of $W$ jets and $B$ QCD jets. We did this for different jet $p_T$' bins (in $50\gev$ steps) separately from $200\gev$ to $1 \tev$ and found a factor of $\sim2$ improvement in $S/\sqrt{B}$ for all $p_T$'s considered. 

Despite the success of the jet grooming algorithms, a good amount of QCD background jets still remain. This is because, although rare, hard splitting does happen in QCD jets which can mimic a $W$ jet more closely. In order to further eliminate QCD backgrounds, we make use of the difference in radiation patterns between $W$ jets and QCD jets, as a result of their different color structures. $W$ is a color singlet particle, therefore, when highly boosted, (almost) all radiation from its decay is contained in a small cone around the $W$ momentum. This is different from a QCD jet, which is initiated from a colored particle with radiation more diffusely distributed. This difference is visible in jet shape variables such as planar flow \cite{planarflow} and N-subjettiness \cite{nsubjettiness}. In Ref.~\cite{wtag}, we proposed a set of simple but powerful variables which we dubbed mass and $p_T$ $R-\cores$. These mass ($p_T$) $R-\cores$ are defined as the ratios between the original jet mass ($p_T$) and the leading subjet mass ($p_T$) reclustered with a set of smaller $R$'s. The $R-\cores$ measure how the masses or $p_T$'s grow with an increasing $R$, which are sensitive to how the radiation is distributed. 

None of the above variables alone can account for all differences between $W$ jets and QCD jets, therefore, the most powerful way to use the variables is to combine them in a multivariate tagging algorithm. In Ref.~\cite{wtag}, we selected 25 most useful variables and combined them using the Boosted Decision Trees method. These variables include the masses and $p_T$'s after jet grooming, planar flows, $p_T$ $R-\cores$, etc.. The multivariate method was applied to jet samples that have passed the filtered mass window cut and another factor of $\sim 2$ improvement is achieved in $S/\sqrt{B}$ on top of the jet grooming algorithms. One may also simplify the method by including fewer variables and achieve improvement nearly as good. For example, we selected a set of 7 variables which give a result about 25\% worse than using the full set. One may also include other variables such as N-subjettiness \cite{nsubjettiness}, charged particle multiplicity \cite{tracking} and volatility \cite{qjet}. However, adding more variables are not likely to significantly improve the $W$ tagging performance because they contain redundant information (e.g. we have checked by including N-subjettiness and charged particle multiplicity). In this article, we use the method in Ref.~\cite{wtag} with the full set of 25 variables.

\end{document}